\documentclass[11pt,twoside]{article}
\usepackage{asp2010}

\resetcounters

\begin{document}

\title{Time-resolved spectroscopy of DROXO X-ray sources: Flares and Fe\,K$\alpha$ emission}
\author{Beate~Stelzer$^1$, Ettore~Flaccomio$^1$, Ignazio~Pillitteri$^2$, Costanza~Argiroffi$^{3,1}$ and Salvatore~Sciortino$^1$
\affil{$^1$ INAF - Osservatorio Astronomico di Palermo, Italy} 
\affil{$^2$ SAO - Harvard Center for Astrophysics, USA}
\affil{$^3$ Dipartimento di Scienze Fisiche ed Astronomiche, Universit\`a di Palermo, Italy}}

\begin{abstract}
We present a systematic search for Fe\,K$\alpha$ emission from young stellar
objects of the $\rho$\,Ophiuchi star forming region 
observed in the {\em Deep Rho Ophiuchi XMM-Newton Observation}. 
\end{abstract}

\section{The {\em Deep Rho Ophiuchi XMM-Newton Observation} (DROXO)}

The {\em Deep Rho Ophiuchi XMM-Newton Observation}, DROXO, 
is an {\em XMM-Newton} Large Project with a nominal exposure time of 
$500$\,ksec centered on core\,F of the $\rho$\,Oph star forming region. 
A catalog of $111$ X-ray sources, more than $50$\,\% of which identified 
with known cloud members, was presented by \cite{Pillitteri10.1}. 
That paper also comprises a detailed description of the general data 
analysis steps
including data filtering, source detection, treatment of the background etc.
Here we present the results of a systematic search for Fe\,K$\alpha$ emission
in all $111$ DROXO X-ray sources. The Fe\,K$\alpha$ emission of one DROXO
source, the Class\,I protostar Elias\,29, has already been discussed 
in a separate paper \citep{Giardino07.2}. Our analysis also includes
this object and we confirm the results by Giardino et al.

\section{Fe\,K$\alpha$ emission in Young Stellar Objects}

The X-ray spectra of Young Stellar Objects (YSOs) can be explained by 
thermal emission from their coronae that are a few times $10^6$\,K hot. 
In roughly a dozen YSOs an emission line at $6.4$\,keV has been identified 
that can not be attributed to highly ionized coronal material 
\citep[e.g.][]{Imanishi01.1, Tsujimoto05.1}.
  In the Sun this line represents K$\alpha$ fluorescence of neutral iron in 
the photosphere irradiated with X-rays from the solar corona during flares. 
In YSOs this line is usually attributed to fluorescence from neutral iron in 
the accretion disk following X-ray illumination from the star's corona. 
In some cases the line equivalent widths 
have been shown to be incompatible with fluorescence calculations 
\citep[e.g.][]{Giardino07.2, Czesla07.2}. 
An alternative excitation mechanism for the line is electron impact ionization.
Giardino et al. suggest the accretion streams between the disk and the stellar
surface as possible location for this process. 
Another possible scenario that may explain the large observed
equivalent widths in these cases without the need to invoke collisional
excitation is the (partial) occultation of the
illuminating continuum flux, e.g. a stellar flare behind the limb
\citep{Drake08.1}. The identification of the origin of the Fe\,K$\alpha$
line in YSOs can greatly 
benefit from time-resolved spectroscopy that enables a direct
evaluation of the continuum visible at the moment when the line is
produced.

\section{Fe\,K$\alpha$ spectroscopy of DROXO X-ray sources}

We have carried out systematic spectral fitting of all $111$ DROXO sources
with XSPEC v.12. 
Spectra were generated for a variety of good-time-intervals, 
spectral binnings, all EPIC detectors individually as well as joint pn+MOS\,1+MOS\,2,  
such that for each X-ray source a total of $24$ time-averaged X-ray spectra
are available. 
Time-resolved spectroscopy is performed in intervals defined using
a maximum likelihood technique that divides the lightcurve in time-intervals of
constant intensity \citep[see e.g.][]{Stelzer07.1}. 

We limit the energy range for spectral fitting to $5-9$\,keV.
At these high energies the photo-absorption, generally an important factor
for YSOs in $\rho$\,Oph 
\citep[median $N_{\rm H}$ for DROXO sources is 
$2.3 \cdot 10^{22}\,{\rm cm^{-2}}$; see ][]{Pillitteri10.1},  
has negligible influence. We adapt
a one-temperature APEC model plus Gaussian feature with fixed energy 
of $6.4$\,keV to the data. 
The $6.4$\,keV feature is considered detected if the normalization of 
the Gaussian is non-zero with $90$\,\% confidence.

\subsection{Results from time-averaged spectra}

\begin{figure}
\begin{center}
\caption{Observed Fe\,K$\alpha$ equivalent widths and X-ray temperature
compared to calculations for photospheric fluorescence by \cite{Drake08.1}.
Solid lines represent models for different heights, $R$, of the exciting 
source in units of stellar radii. Error bars for the data points 
are $68$\,\% confidence levels. 
}
\label{fig:average}
\includegraphics[width=10.0cm, angle=0]{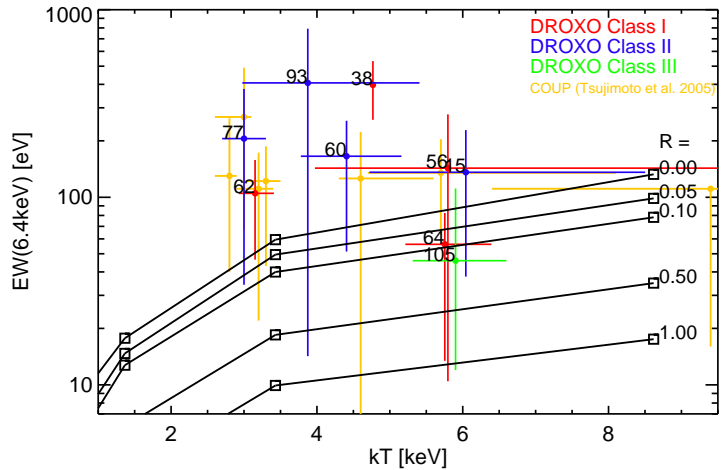}
\end{center}
\end{figure}

The time-averaged spectra include data from $6$ satellite orbits.
The $6.4$\,keV line is detected in $9$ of $111$ DROXO sources: 
$4$\,Class\,I, $4$\,Class\,II, $1$\,Class\,III objects. 

Searching for the origin of the line, we compare the observed line equivalent 
widths to calculations for photospheric fluorescence 
\citep{Drake08.1}. The fluorescence model consists in a single coronal 
flare illuminating the photosphere from a height, $R$. 
The flux in the fluorescence line depends essentially on four parameters:
the iron abundance in the fluorescence medium, the temperature
of the ionizing coronal plasma ($kT$), its height ($R$), and 
the heliocentric angle ($\theta$); \cite{Basko79.1}. 
The models shown in Fig.~\ref{fig:average} are based on the assumption 
of $\theta = 0$, i.e. a flare erupting on the star in the direction of the line-of-sight, 
the case that gives the highest fluorescence flux. 
The calculations for $R=0$ apply also to a fluorescent disk illuminated
by a stellar flare. 

The sample of equivalent widths observed in DROXO is complemented
in Fig.~\ref{fig:average} by
YSOs with Fe\,K$\alpha$ detections from the {\em Chandra Orion Ultradeep
Project} (COUP) discussed by \cite{Tsujimoto05.1}. 
Most observed equivalent widths are compatible with the predictions of the 
photospheric fluorescence model with a flare at moderate height
above the photosphere or a flare irradiating a disk.  
We detect the $6.4$\,keV line from 
one Class\,III object, GY\,463, where material related to a disk 
is not a viable option for the line emission site 
(see also Sect.\ref{subsect:resolved}).

\subsection{Results from time-resolved spectra}\label{subsect:resolved}

\begin{figure}
\begin{center}
\caption{Equivalent widths of $6.4$\,keV emission detected in time-resolved
spectra vs. the time-averaged value. 
Evidently, the time-averaged equivalent width is a single value for a given
star. In the figure, 
for better visibility of all time-resolved measurements for a given star, 
it has been attributed small offsets from its observed value. 
%for the individual data points
%corresponding to different time-intervals for better visibility. 
%the time-averaged value has been slightly offset for the successive data points
}
\label{fig:resolved}
\includegraphics[width=9.5cm, angle=0]{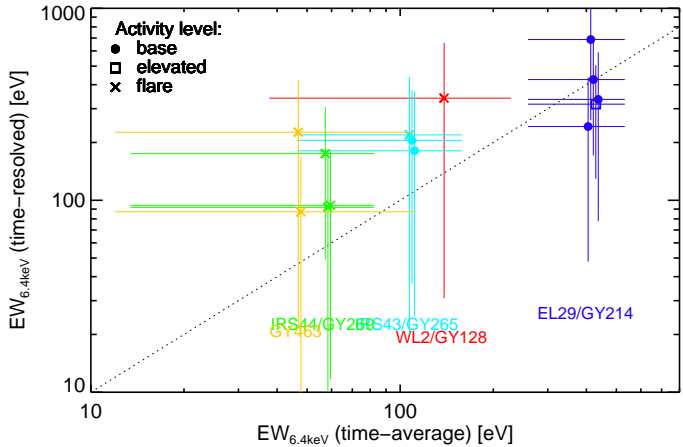}
\end{center}
\end{figure}

There are several reasons to expect the Fe\,K$\alpha$ flux 
to be variable in time. 
Generally, any variation of the ionizing spectrum (in terms of
temperature or emission measure) has an impact on the strength of the
line. Specifically, 
if flares are responsible for its excitation a correlation between 
the presence of Fe\,K$\alpha$ emission and flaring is predicted. 
In practice, the relation between flares and K$\alpha$ emission 
may be concealed by occultation effects as proposed by \cite{Drake08.1}, 
and time-delays between these two phenomena may 
yield geometrical constraints. 

Our time-resolved spectroscopy of DROXO sources results in Fe\,K$\alpha$
detections in individual time segments for $5$ YSOs. 
In Fig.~\ref{fig:resolved} we show the equivalent width of the $6.4$\,keV 
feature in individual time segments 
vs. the equivalent width in the time-averaged spectrum for those objects. 
This comparison demonstrates how  
mixing different activity states by averaging over long time 
intervals yields a smaller equivalent width than the actual maximum value
achieved during the observation. 

\begin{figure}
\begin{center}
\caption{X-ray spectrum with Fe\,K$\alpha$ line fixed at an energy of $6.4$\,keV
for the Class\,III source GY\,463 representing the time-interval with a large flare. 
Red is the coronal one-temperature model, and green the Gaussian representing the
Fe\,K$\alpha$ emission. Line equivalent width and $68$\,\% uncertainties are
written on the top of the figure 
at the line position which is marked with a vertical line.}
\label{fig:gy463}
\includegraphics[width=7.0cm, angle=0]{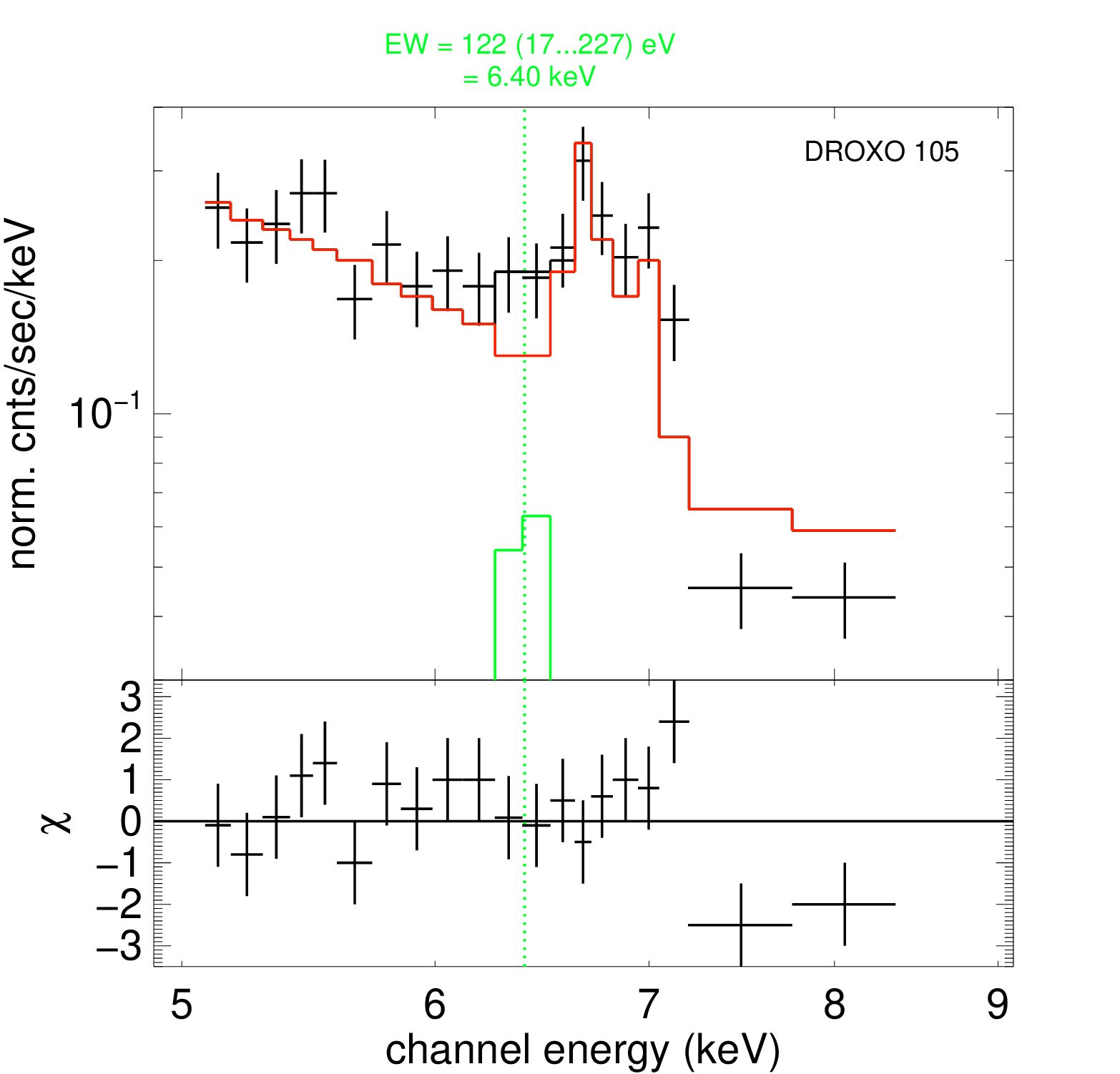}
\end{center}
\end{figure}

In Fig.~\ref{fig:resolved} we distinguish three activity states 
(base level, elevated level and flare) defined by different
intensity in the lightcurve. 
Remarkably, the $6.4$\,keV line is detected not only during flares but 
in some cases during quiescence (`base level'). 
Especially, for Elias\,29 we recover
the result of \cite{Giardino07.2}.
Elias\,29 is also the star with the
highest equivalent width (up to $700$\,\AA). 
It is the only object in 
our sample that is incompatible with the fluorescence model 
taking into account the uncertainties (Source\,38 in Fig.~\ref{fig:average}). 
The second star in which the Fe\,K$\alpha$ emission is observed outside
evident flares is IRS\,43. This source is discussed in more detail
in Sect.~\ref{subsect:irs43}. 
For the only Class\,III object with Fe\,K$\alpha$ emission, GY\,463, 
the time-resolved spectral analysis shows that 
the detection of this line is related to a large flare identified
in the lightcurve.   
The spectrum corresponding to the time-interval where the line was detected 
is shown in Fig.~\ref{fig:gy463}.

\subsection{The puzzling Fe\,K$\alpha$ emission of IRS\,43}\label{subsect:irs43}

\begin{figure}[t]
\begin{center}
\parbox{14.cm}{
\parbox{7.0cm}{
\caption{Three different spectra around the $6.4$\,keV region for IRS\,43: 
(top right) time-averaged with fixed line position, (bottom left) time-averaged with free line position, and (bottom right) restricted to the time-interval during which the line was detected in the time-resolved spectroscopy; see Sect.~\ref{subsect:irs43} for details and caption of Fig.~\ref{fig:gy463} for an explanation of the features shown in the figure.}
\label{fig:irs43}
}
\parbox{0.5cm}{\hspace*{0.5cm}}
\parbox{7.0cm}{
\includegraphics[width=7.0cm, angle=0]{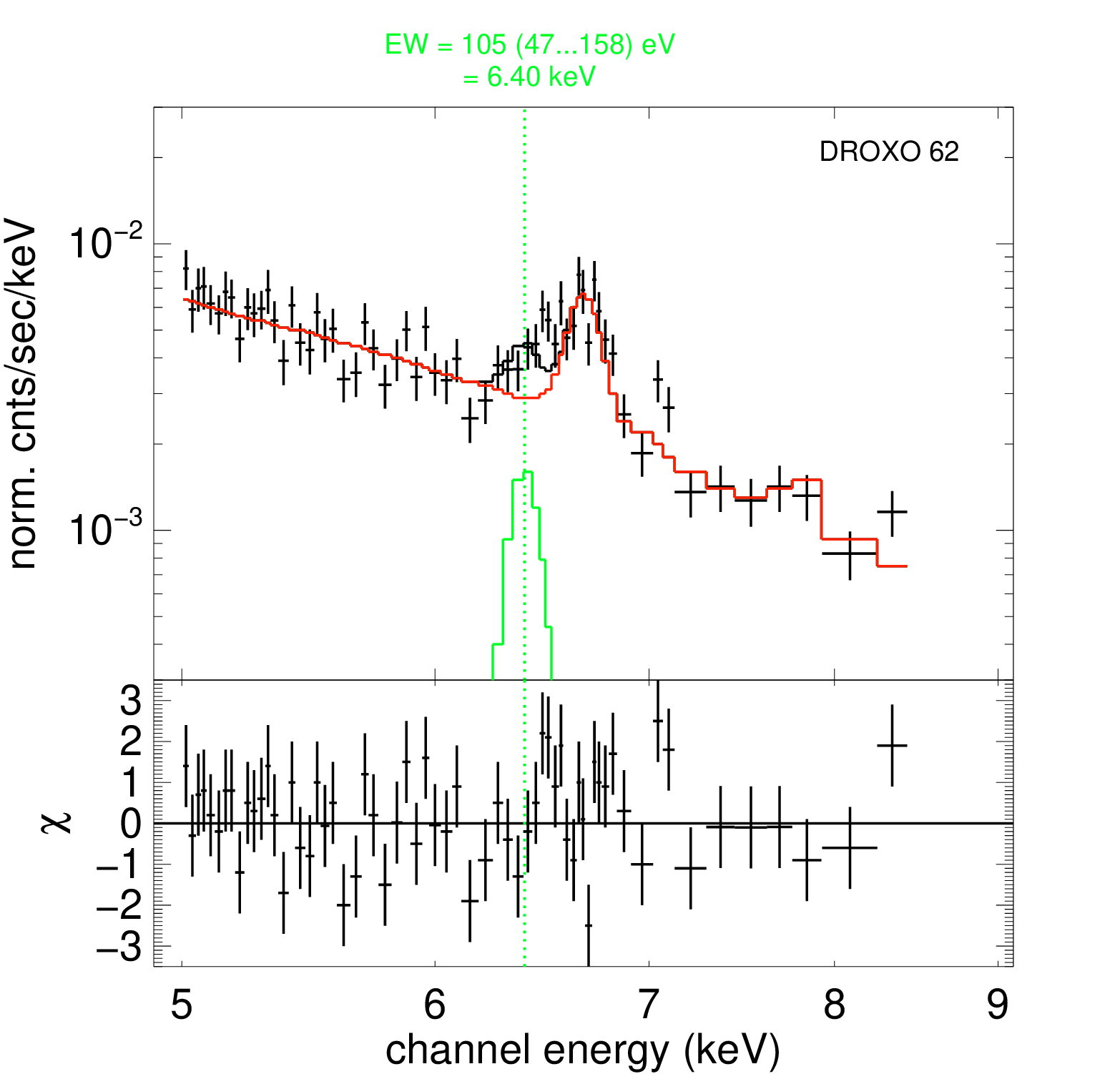}
}
}
\parbox{14.cm}{
\parbox{7.0cm}{
\includegraphics[width=7.0cm, angle=0]{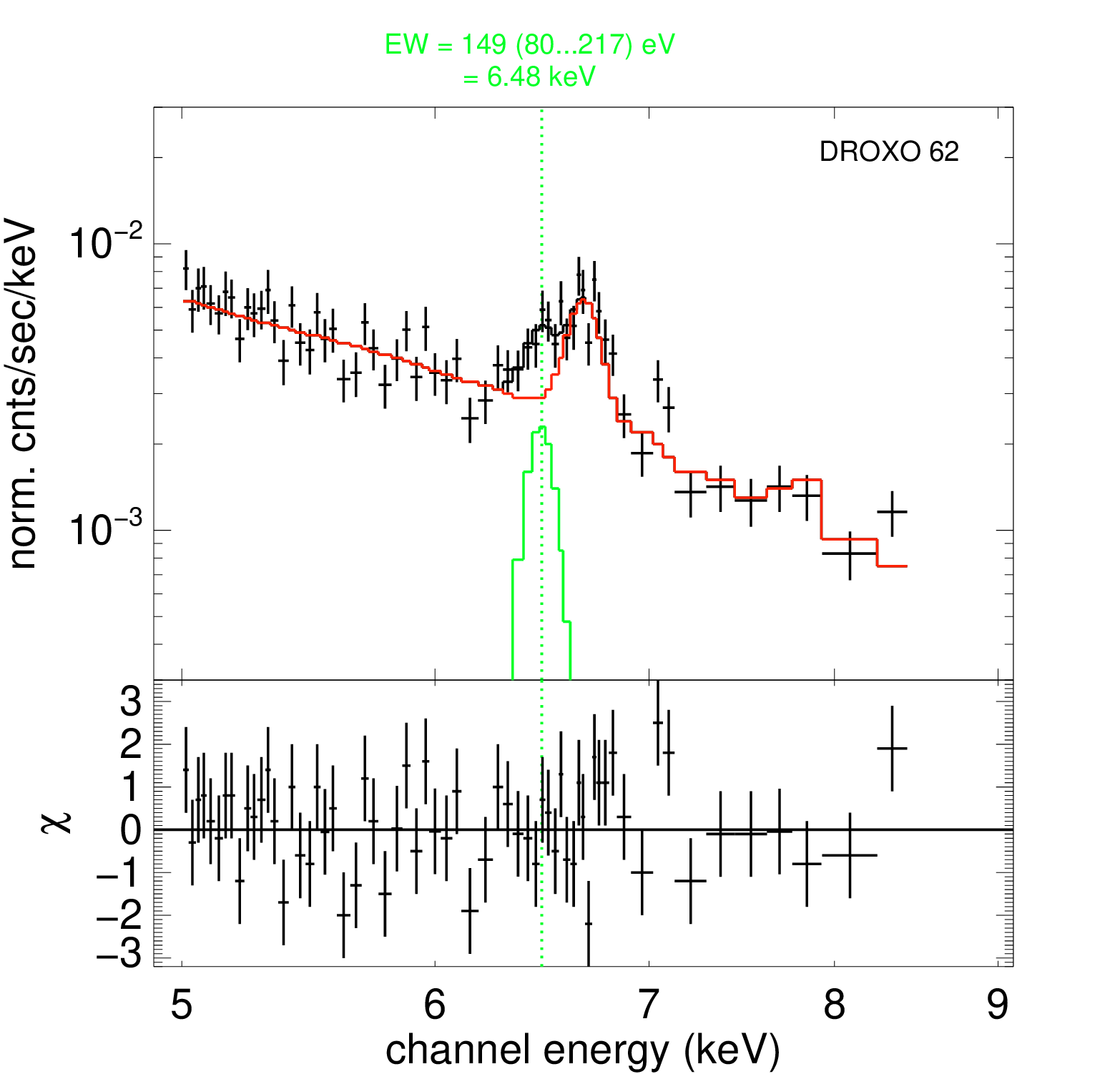}
}
\parbox{7.0cm}{
\includegraphics[width=7.0cm, angle=0]{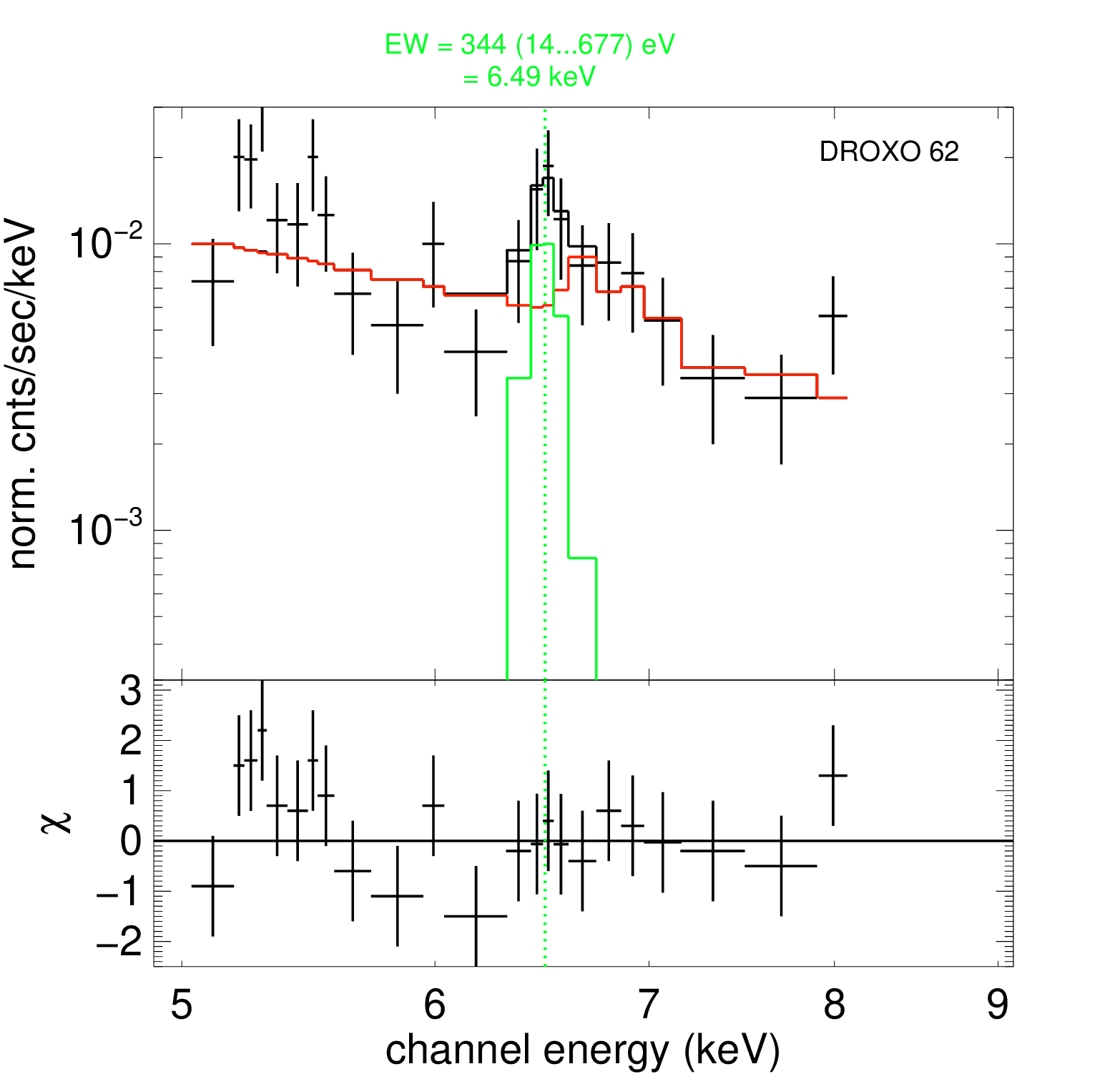}
}
}
\end{center}
\end{figure}

The Fe\,K$\alpha$ emission is located at $6.4$\,keV for neutral and low
ionized iron with very little dependence of the fluorescence energy
on the charge state \citep{Kallman04.1}. 
 This justifies our choice of fixing the line position
in the spectral fitting process. However, in the time-averaged spectrum of 
%the Class\,I object 
IRS\,43 (alias YLW\,15) 
this approach results in a clear residual between the model and the data
that stems from an excess flux at energies between the Gaussian (fixed
at $6.4$\,keV) and the coronal iron line at $6.7$\,keV; 
see Fig.~\ref{fig:irs43} top right. 

IRS\,43 is a binary with $0.5^{\prime\prime}$ separation. The 
northern component is identified only at radio
and mid-IR wavelengths, i.e. it is probably either a very young protostar
or a knot in the outflow \citep[e.g.][]{Girart00.1, Haisch02.1}.
The X-ray emission is most likely produced on the brighter object, 
IRS\,43\,S, a Class\,I protostar according to the (unresolved) 
spectral energy distribution. \cite{Greene02.1} detected stellar
absorption features in near-IR spectra of IRS\,43 that point to a 
spectral type of K5\,IV/V. From the near-IR photometry they 
derive an extinction of $A_{\rm V} \sim 40$\,mag and 
a stellar luminosity of $\sim 3\,L_\odot$ leaving $70$\,\% of its
bolometric luminosity \citep[$10\,L_\odot$; ][]{Wilking89.1} to the accretion
process. However, the conclusion on the accretion luminosity 
relies on the interpretation of the second
radio and mid-IR source as a jet and not as an embedded binary companion that
would contribute to the luminosity derived from the spectral energy
distribution. 
 
A fit of the DROXO spectrum 
with free line position results in a line energy of $6.48 \pm 0.05$\,keV
where the uncertainty is for a $68$\,\% confidence level 
(Fig.~\ref{fig:irs43} bottom left). 
This line is apparently produced by significantly ionized ions 
(Fe\,XIX -- Fe\,XXIII). 
Those ions are dominating the ionization balance of a collisional plasma 
in a small temperature range of $8-14$\,MK. 
It is therefore difficult to attribute 
this line to fluorescence from a photosphere or an accretion disk where such
high temperatures are not reached. It seems like in this case 
the fluorescence takes place in the corona itself. 
However, the coronal column density is probably too small to produce the
observed equivalent width of $\sim 150$\,eV. 
Another hypothesis is that this fluorescence line is emitted in denser plasma 
such as the one present in the chromosphere or transition region. 
The emitting material may then be photo-ionized rather than collisionally ionized, 
and this would allow for lower temperature. 
Comparison of the observed line energy with the high-resolution profiles 
for a photo-ionized plasma calculated by 
\cite{Kallman04.1} suggests an ionization parameter of $\log{\zeta} \sim 2$.
For this value of $\zeta$, the temperature of the photoionization
calculations is $\log{T}\,{\rm [K]} \sim 5.2$ 
and the Fe\,K$\alpha$ line emissivity is dominated by Fe\,XIX. 
%Note that, the emissivity of a given ion has its maximum at the ionization parameter
%where its parent ion dominates the ionization balance. 

%Surprisingly, the lightcurve
%does not hint at any particular event, such as a flare, related with the occurrence
%of this ion fluorescence. 
Inspection of the lightcurve and the time-resolved spectroscopy for IRS\,43
reveals that the Fe\,K$\alpha$ line is detected during $4$ 
time-segments. Only one of them shows the energy shift. 
This time-interval corresponds to a small flare during the last orbit
of the observation. The spectrum of this time segment is shown in 
the bottom right panel of Fig.~\ref{fig:irs43}. 
One of the remaining three time-intervals with
Fe\,K$\alpha$ emission has a larger flare, while the other two represent
the quiescent X-ray state. A possible scenario for the shifted line energy
may be fluorescence from the chromosphere or transition region illuminated by
a flare occuring near or partially behind the limb.

\acknowledgements 
We acknowledge the support of the ASI/INAF agreement \newline
I/099/10/0.

\bibliography{stelzer_b}

\end{document}